\documentclass[sigconf]{acmart}
\PassOptionsToPackage{hyphens}{url}
\PassOptionsToPackage{breaklinks}{hyperref}
\expandafter\def\expandafter\UrlBreaks\expandafter{\UrlBreaks
  \do\a\do\b\do\c\do\d\do\e\do\f\do\g\do\h\do\i\do\j%
  \do\k\do\l\do\m\do\n\do\o\do\p\do\q\do\r\do\s\do\t%
  \do\u\do\v\do\w\do\x\do\y\do\z\do\A\do\B\do\C\do\D%
  \do\E\do\F\do\G\do\H\do\I\do\J\do\K\do\L\do\M\do\N%
  \do\O\do\P\do\Q\do\R\do\S\do\T\do\U\do\V\do\W\do\X%
  \do\Y\do\Z}
\def\BibTeX{{\rm B\kern-.05em{\sc i\kern-.025em b}\kern-.08emT\kern-.1667em\lower.7ex\hbox{E}\kern-.125emX}}

\usepackage{longtable}
\usepackage{soul}
\usepackage{colortbl}
\usepackage[hyphens]{url}
\usepackage[breaklinks]{hyperref}
\usepackage{subfigure}
\usepackage{booktabs}
\usepackage{tabularx}
\usepackage{graphicx}
\usepackage{algpseudocode}
\usepackage{amsfonts}
\usepackage{amsmath}
\usepackage{amssymb}
\usepackage{color}
\usepackage{comment}
\usepackage[ruled,vlined,linesnumbered,procnumbered]{algorithm2e}
\usepackage{balance}

\SetAlgoCaptionSeparator{}

\definecolor{Gray}{gray}{0.85}

\makeatletter
\newcommand{\colvect}[2][r]{%
  \gdef\@VORNE{1}
  \left(\hskip-\arraycolsep%
    \begin{array}{#1}\vekSp@lten{#2}\end{array}%
  \hskip-\arraycolsep\right)}

\def\vekSp@lten#1{\xvekSp@lten#1;vekL@stLine;}
\def\vekL@stLine{vekL@stLine}
\def\xvekSp@lten#1;{\def\temp{#1}%
  \ifx\temp\vekL@stLine
  \else
    \ifnum\@VORNE=1\gdef\@VORNE{0}
    \else\@arraycr\fi%
    #1%
    \expandafter\xvekSp@lten
  \fi}

\copyrightyear{2019}
\acmYear{2019}
\setcopyright{acmlicensed}
\acmConference[WebSci 2019]{ACM Web Science Conference 2019}{June 30--July 3, 2019}{Boston, MA, USA}
\acmPrice{}
\acmDOI{}
\acmISBN{}

\begin{document}

\title[Better Safe Than Sorry]{Better Safe Than Sorry:\\an Adversarial Approach to improve Social Bot Detection}
\author{\mbox{Stefano Cresci, Marinella Petrocchi}}
\affiliation{%
  \institution{IIT-CNR, Pisa, Italy}
}
\email{[name.surname]@iit.cnr.it}


\author{Angelo Spognardi}
\affiliation{
  \institution{Dept. of Computer Science,\\Sapienza University of Rome, Italy}
}
\email{spognardi@di.uniroma1.it}

\author{Stefano Tognazzi}
\affiliation{%
  \institution{IMT School for Advanced Studies Lucca, Italy}
}
\email{stefano.tognazzi@imtlucca.it}

\renewcommand{\shortauthors}{S. Cresci \textit{et al.}}

\begin{abstract}
  The arm race between spambots and spambot-detectors is made of
  several cycles (or generations): a new wave of spambots is created
  (and new spam is spread), new spambot filters are derived and old
  spambots mutate (or \textit{evolve)} to new species.
  Recently, with the diffusion of the adversarial
  learning approach, a new practice is emerging: to manipulate on
  purpose target samples in order to make stronger detection
  models. Here, we manipulate generations of Twitter
  social bots, to obtain - and study - their possible future evolutions, with the  aim
 of  eventually deriving more effective detection techniques. In detail, we
  propose and experiment with a novel genetic algorithm for the
  synthesis of online accounts. The algorithm allows to create
  synthetic \textit{evolved} versions of current state-of-the-art
  social bots. Results demonstrate that synthetic bots really escape
  current detection techniques. However, they give all the needed
  elements to improve such techniques, making possible a
  proactive approach for the design of social bot detection systems.
\end{abstract}


\maketitle


\section{Introduction}
\label{sec:intro}

A worrying peculiarity of spammers and bots (or spambots) is that they \textit{evolve} over time, adopting sophisticated techniques to evade well-established detection systems~\cite{yang2013,Cresci2017}.
In the context of Online Social Networks (ONSs), newer social spambots often feature advanced characteristics that make them way harder to detect with respect to older ones, since capable of mimicking human behaviors and interaction patterns better than ever before~\cite{ferrara2016,Cresci2017}. 
These automated accounts represent -- to the best of the literature knowledge -- the third and most novel generation of social bots, following the original wave dated back in the 00s, and passing through a second generation dated around 2011~\cite{yang2013}. 
The latest social bots are capable of sharing (credible) fake news, inflating the popularity of OSN users, and reshaping political debates~\cite{websci18,steward2018}. Given this picture, it is not surprising that evolution mechanisms (together with coordination and synchronization ones)
 represent one of the key factors that currently allow malicious accounts to massively tamper with our social ecosystems~\cite{cresci2018fake}.

Despite malicious accounts evolution representing a sort of Pandora's box, little to no attention has been posed towards studying -- and possibly anticipating -- such evolution.
In fact, in past years, as social bots gradually became clever in escaping detection, scholars and OSNs administrators tried to keep pace (i.e., \textit{reacted}), by proposing ever more complex detection techniques, as described in the Related Work section. 
The natural consequence of this \textit{reactive} approach -- according to which a new technique is designed only after having collected evidence of  new mischiefs of evolved bots -- is that researchers and OSN admins are constantly one step behind the bot developers~\cite{cresci2018proaction}. 

The classification task of recognising if an online account is genuine or not is {\it adversarial} in nature, being focused on distinguishing {\it bad}
  samples from {\it good} ones. Nonetheless, a new approach is gaining momentum in the wide field of artificial intelligence, leveraging the concept of {\it adversarial learning}: the automatic learning within a hostile environment~\cite{kurakin2017adversarial}. This allows to both discover vulnerabilities in learning algorithms and to test algorithmic techniques which yield more robust
  learning~\cite{DBLP:journals/corr/abs-1708-08327}.												
  Intuitively,  the evolution of new waves of social bots can be seen
  as a problem of {\it adversarial classifier evasion}, where the attacker
  changes the generated samples to evade detection. Thus, the core idea of this paper  is
  to manipulate \textit{on purpose} target samples in order
  to produce stronger detection models.
  Inspired by the adversarial learning approach, for the first time to the
  best of our knowledge, we carry out an exploratory investigation to
  define and implement a \textit{proactive} technique to  study and
  detect evolving social bots.

Specifically, we aim at answering the
following critical, yet unexplored, research questions:

\textbf{RQ1 --} \textit{Can we develop an analytical framework for simulating spambot evolutions?}

\textbf{RQ2 --} \textit{Can we use such framework for synthesizing new generations of spambots? And, most importantly, are these evolved spambots capable of going undetected by state-of-the-art techniques?}

\textbf{RQ3 --} \textit{Can we leverage this proactive and adversarial study of spambot evolutions to improve current detection techniques?}

\vspace{.5em}
\noindent \textbf{Approach.} Our methodological approach to answer the three questions stems from the so-called \textit{digital DNA} technique~\cite{cresci2016dna}, where the behavioral lifetime of an account is encoded as a sequence of characters, built according to the chronological sequence of actions performed by the account.
The adoption of this behavioral modeling technique offers a DNA-like representation for the lifetime of each account, including new social spambots. We feed the DNA sequences to a custom-designed genetic algorithm, so as to study possible evolutions of the accounts~\cite{Mitchell:1998}. 
The customized genetic algorithm iteratively selects the \textit{best} evolutions, so as to converge towards synthetic bots capable of resembling behavioral characteristics of legitimate accounts.
Furthermore, by constraining possible evolutions within the algorithm, the approach allows to obtain synthetic accounts capable of performing specific tasks (e.g., viral marketing, message spamming, mass retweeting, etc.).
Notably, the digital DNA behavioral modeling technique has been exploited in the past as the building block of a state-of-the-art detection system~\cite{cresci2017social}. Thus, we can also apply such detection technique on 
the synthetically evolved accounts,  to evaluate whether they are capable of evading detection.

\vspace{.5em}
\noindent \textbf{Contributions.} This work contributes along several dimensions. Firstly, (i) we propose \textsc{GenBot}, a novel genetic algorithm specifically designed for social spambot evolutions.
 By employing a cost function that quantifies the difference between a new generation of spambots and a group of legitimate accounts, \textsc{GenBot} is capable of generating spambots 
 whose behavior is similar to that of the legitimate ones. Notably, our results outperform previous attempts to simulate the behavior of human accounts~\cite{dsaa2017}.
 Then, (ii) we discuss the design of an analytical framework for simulating possible social spambot evolutions that leverages both the digital DNA behavioral modeling technique and the genetic algorithm previously defined. We experiment with this framework to synthesize a novel generation of evolved spambots, with the aim of producing adversarial samples.
 Furthermore, (iii) we assess the extent to which the newly synthesized social bots are detected by 3 state-of-the-art techniques. Results show that, with the proposed framework, it is possible to create an adversarial behavioral fingerprint that allow bots to escape detection.
 Finally, (iv) by studying the characteristics of the synthetic evolved spambots, we draw useful insights into which account features could be considered in order to improve the detection of real evolving spambots.

\vspace{.5em}
\noindent \textbf{Broadening the approach.}
We ground our study on a recently-proposed proactive approach to spambot detection~\cite{cresci2018proaction}. In order to carry out extensive experimentation on real-world data, without loss of generality, here we implement it by focusing on the behavior (i.e., the sequences of actions that accounts perform) of spambots and legitimate accounts. This choice opens up the possibility to leverage, for our experiments, the \textit{digital DNA} behavioral modeling technique~\cite{cresci2016dna} as well as the \textit{social fingerprinting} spambot detection technique~\cite{cresci2017social}. However, despite this particular implementation of the proactive approach, similar analyses could be carried out, by relying on different modeling and spambot detection techniques, such as those based on network/graph analysis and those based on content analysis.

\vspace{.5em}
\noindent \textbf{Reproducibility.} Both the data\footnote{\url{http://mib.projects.iit.cnr.it/dataset.html}} and the code\footnote{\url{http://sysma.imtlucca.it/tools/digdna-genetic-algorithm/}} used in this study are publicly available for scientific purposes.

\section{Background and notation}
\label{sec:background}
Here, we provide a succinct description of the main concepts related to digital DNA sequences and genetic algorithms, as well as notations and metrics adopted in the remainder of this study.

\subsection{Digital DNA}
\label{sec:digital-dna}

\newcommand{\Bttype}{\ensuremath{\mathbb{B}}}

\begin{figure}[t]
  \centering
  \includegraphics[width=0.85\columnwidth]{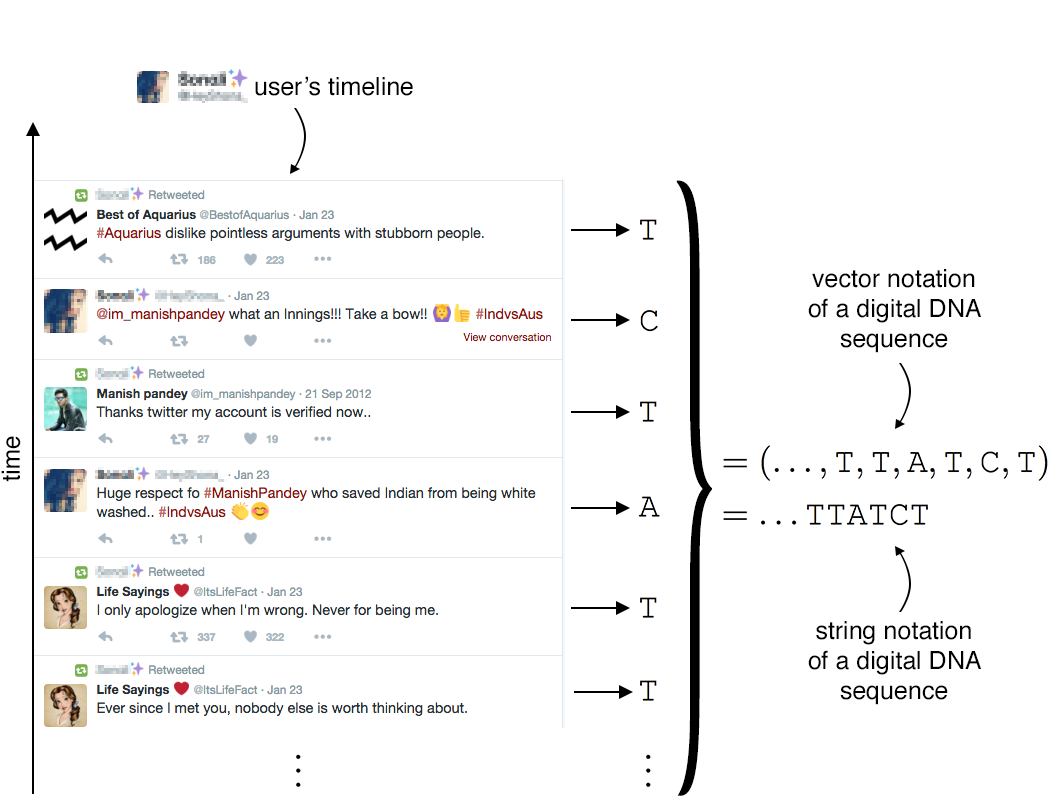}
  \caption{Excerpt of a digital DNA extraction process for a Twitter user with the alphabet $\mathbb{B}=\{$\texttt{A,C,T}$\}$, where \texttt{T} is assigned to every tweet, \texttt{C} to every reply, and \texttt{A} to every retweet. 
  \label{fig:timeline}}
\end{figure}

\noindent \textbf{Digital DNA sequences.} We define a digital DNA sequence $\mathbf{s}$ as a row-vector of characters (i.e., a string),
\begin{equation*}
\mathbf{s} = (b_1, b_2, \ldots, b_n) \quad b_i \in \mathbb{B} \;\; \forall \;\; i = 1, \ldots, n
\end{equation*}
Characters $b_i$ in $\mathbf{s}$ are also called the (DNA) \textit{bases} and are drawn from a finite set  $\mathbb{B}$, called \textit{alphabet},
\begin{equation*}
\label{eq:alphabet-def}
\mathbb{B} = \{B_1, B_2, \ldots, B_N\} \quad \forall\; i \neq j: B_i \neq B_j \;\; 
\end{equation*}
Online users' behaviors can be represented by encoding each user
action, in chronological order, with an appropriate base. In this way,
we obtain the sequence of characters that makes up the digital DNA
sequence of the user. For example, Figure~\ref{fig:timeline} shows the
process of extracting the digital DNA sequence of a Twitter user, by
scanning its timeline according to the alphabet
$\mathbb{B} = \{\verb|A|, \verb|C|, \verb|T| \}$, 
where \verb|T| is assigned to every tweet, \verb|C| to every reply,
and \verb|A| to every retweet.
A digital DNA sequence can be represented, then, with a compact string
like $\mathbf{s}=\ldots\verb|TTATCT|$.
%
%
Additional details about the theoretical foundations of digital DNA can be found in~\cite{cresci2016dna,cresci2017social}. 

\vspace{.5em}
\noindent \textbf{Similarity between digital DNA sequences.} In order to analyze groups of users rather than single users, we need to study multiple digital DNA sequences as a whole.  A group $\mathbf{A}$ of $M = | \mathbf{A} |$ users can be described by the strings representing the digital DNA sequences of the $M$ users.

To perform our analyses on digital DNA sequences, we can rely on
recent advances in the fields of bio-informatics and string mining~\cite{gusfield1997}.
One of the possible means to quantify
similarities between sequential data representations 
is the \textit{longest common
  substring}~\cite{arnold2011}.  
Given two strings,
$\mathbf{s}_i$ of length $n$ and $\mathbf{s}_j$ of length $m$, their
longest common substring (henceforth LCS) is the longest string that
is a substring of both $\mathbf{s}_i$ and $\mathbf{s}_j$. For example,
given $\mathbf{s}_i = \verb|MASSACHUSETTS|$ and
$\mathbf{s}_j = \verb|PARACHUTE|$, their LCS is the string \verb|ACHU|
and the LCS length is 4.  The extended version of this problem, which
considers an arbitrary finite number of strings, is called the
\textit{k-common substring} problem~\cite{chi1992}. In this case,
given a vector $\mathbf{A} = (\mathbf{s}_1, \dots, \mathbf{s}_M)$ of
$M$ strings, the problem is that of finding the LCS that is common to
at least $k$ of these strings, for each $2 \le k \le M$.  Notably,
both the \textit{longest common substring} and the \textit{k-common
  substring} problems can be solved in linear time and space, by
resorting to the generalized suffix tree and by implementing
state-of-the-art algorithms, such as those proposed
in~\cite{arnold2011}. Given that, in the \textit{k-common substring}
problem, the LCS is computed for each $2 \le k \le M$, it is possible
to plot a \textit{LCS curve}, showing the relationship between the
length of the LCS and the number $k$ of strings~\cite{cresci2017social}.

\begin{figure}
  \centering
  {\includegraphics[width=0.25\textwidth]{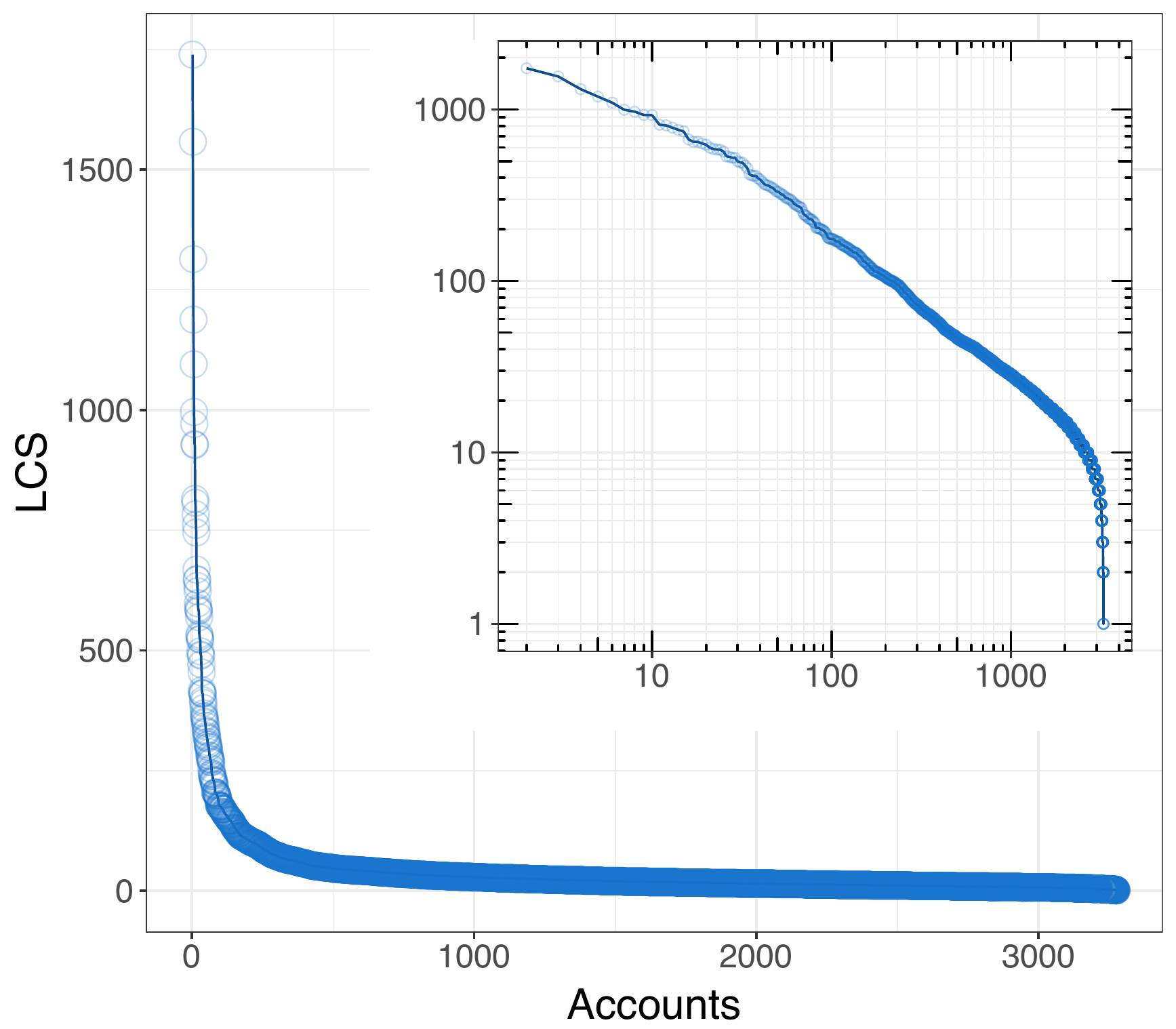}}
  \caption{LCS curve of a group of legitimate (human-operated) accounts.
  \label{fig:lcs-tweettype3-human}}
\end{figure}

Figure~\ref{fig:lcs-tweettype3-human} depicts the LCS curve computed for a group of legitimate (human-operated) Twitter accounts, via the alphabet $\mathbb{B} = \{\verb|A|, \verb|C|, \verb|T| \}$. 
On the $x$ axis is reported the number $k$ of accounts (corresponding to the $k$ digital DNA sequences used to compute LCS values) and on the $y$ axis the length of the LCS common to at least $k$ accounts. Therefore, each point in a LCS curve corresponds to a subset of $k$ accounts that share the longest substring (of length $y$) among all those shared between all the other possible subsets of $k$ accounts.

A LCS curve is a representation of the behavioral similarities among a group of users, since it is an ordered sequence of substring lengths. To obtain a single value as \textit{measure} of similarity for the whole group, we can compute the area under the LCS curve (AUC)~\cite{friedman2001,fawcett2006}. Since LCS curves are discrete functions defined over the $[2, M]$ range, their AUC can be computed straightaway, without approximations, with the following trapezoid rule,
\begin{eqnarray}
\label{eq:sequence-def}
\text{AUC} & = & \sum_{k = 3}^{M}{\frac{(\text{LCS}[k-1]+\text{LCS}[k])\Delta_k}{2}} 
\end{eqnarray}
Compared to LCS, the definition of AUC, given in Equation~(\ref{eq:sequence-def}), allows to quantitatively and directly compare the overall behavioral similarity among different groups.
This notion of AUC is exploited in Section~\ref{sec:results} to evaluate the results of this study.

\subsection{Genetic algorithms}
\label{sec:genetic-algo}

\SetAlFnt{\footnotesize\sf}
Genetic algorithms~\cite{Mitchell:1998} represent a popular meta-heuristic technique to solve optimization problems. It is inspired by the natural evolution, in which only the best candidates of some species survive across several subsequent generations. 
Starting from a random sample of candidate solutions (a so-called \emph{population} of \emph{individuals}), only the best ones are elected to evolve.
Similarly to what happens in nature, through a mechanism of recombination and mutation of those individuals, a new \emph{generation} is obtained. This evolved generation is expected to have a better quality with respect to the previous one.
%
The quality of individuals is evaluated by associating a \emph{fitness score} to all of them. The score is computed by a given \emph{fitness function} that is designed according to the task at hand.
During each generation, the current population is modified via either a single input function called \emph{mutation} or a two input function called \emph{crossover}. The outputs of these functions are called \emph{offsprings} and a new generation is formed by merging some of the individuals from the previous generation with some of the offsprings. In our experimental scenario, when we refer to an individual, we refer to a group of users. Thus, our population is composed of different groups of users that evolve passing from a generation to the next one.

\vspace{.5em}
\noindent \textbf{Notations.} We call $P_0$ the initial population and $P_i$ the population at the $i$-th generation.
Given a population $P_i$, $G_j$ = $\{u_1, u_2, \ldots, u_M \}$ is the $j$-th individual (i.e., a group of users) of $P_i$.
Then, $G_j[k] = u_k$ is the $k$-th user of the group $G_j$, characterized by its digital DNA sequence. In other words, $u_k$ is a digital DNA sequence encoding the behavior of the $k$-th Twitter user.
We define $v_j$ as the fitness score of the $j$-th individual. The population $P_i$ = $\{(G_1,v_1),(G_2,v_2),\ldots,(G_J,v_J)\}$
is a set of pairs containing the individuals and their associated fitness scores.



\section{An algorithm for simulating bot evolutions}
\label{sec:our-algo}
The design of a custom genetic algorithm for solving a given task involves the definition of the parameters and functions used by the algorithm in its iterative execution. In this section, we first describe the design choices and building blocks of the \textsc{GenBot} algorithm and we conclude by defining the algorithm itself.

\subsection{Building blocks: fitness, mutation and crossover}
\noindent \textbf{Fitness.} In our scenario, individuals of best quality are groups of bots that best emulate the behavior of a group of legitimate (human-operated) accounts. 
To formalize this intuition, we rely on the notion of behavioral similarity expressed by digital DNA and LCS curves.
More specifically, our goal for this task is to \emph{minimize} the distance between the LCS curve (i.e., the behavioral representation) of a group of legitimate accounts and the LCS curve of a population of synthetic evolved bots.

We rely on the Kullback-Liebler distance ($D_{KL}$) to compute the distance between two LCS curves, since it has already been fruitfully employed in recent similar work~\cite{viswanath2015,dsaa2017}. $D_{KL}$ is an information theoretic metric that measures how much information is lost when a target probability distribution ${P}_X(x)$ is approximated by $\hat{P}_X(x)$. In detail, $D_{KL}$ is the symmetric version of the Kullback-Liebler divergence $d_{KL}$ (defined as,
$d_{KL}(\hat{P}_X, {P}_X) = \sum_x{\ln\left(\frac{\hat{P}_X(x)}{P_X(x)}\right)\hat{P}_X(x)}$),
%
where,
\begin{equation}
\label{eq:kl-distance}
D_{KL}(\hat{P}_X, P_X) = \frac{d_{KL}(P_X, \hat{P}_X)+d_{KL}(\hat{P}_X,P_X)}{2}
\end{equation}

In the \texttt{GenBot} algorithm, the target distribution ${P}_X(x)$ is obtained from the LCS curve of legitimate accounts, while the approximating distribution $\hat{P}_X(x)$ is obtained from the LCS curve of a group of bots. 
The $D_{KL}$ distance is computed by the fitness function \texttt{Fit}$(\cdot)$ that accepts as input an individual $G_j$ and the target LCS curve $b$ of legitimate accounts.
The output is a scalar $v_j$ that represents the fitness score of the individual $G_j$. 


\SetKwFunction{rand}{rand}
\DontPrintSemicolon
\makeatletter
\newcommand{\removelatexerror}{\let\@latex@error\@gobble}
\makeatother

\begin{figure}
	\begin{minipage}[t]{0.23\textwidth}
		\removelatexerror
		\vspace{0pt}
		\hspace{-0.4cm}
		\begin{algorithm}[H]
			\scriptsize
			\NoCaptionOfAlgo
			
			\caption{\texttt{Fit}($\mathit{G_j,b}$)\label{alg:fit}}
			$g \gets LCS(G_j)$ \;
			$v_j \gets D_{KL}(g,b)$\;
			\KwRet{$v_j$}\;
		\end{algorithm}
	\end{minipage}
	\hspace{-0.2cm}
	\begin{minipage}[t]{0.23\textwidth}
		\removelatexerror
		\vspace{0pt}
		\begin{algorithm}[H]
			\scriptsize
			\NoCaptionOfAlgo
			
			\caption{\texttt{GCO}($\mathit{G_x , G_y, r }$)\label{alg:outCx}}
			\For{ $ i \in \{1,\dots , r\} $}{
			    $G_{xy}[i] \gets G_{x}[i]$\;
			    $G_{yx}[i] \gets G_{y}[i]$\;
			    }
			\For{ $ i \in \{r+1, \dots , |G_x| \} $ }{
			    $G_{xy}[i] \gets G_{y}[i]$\;
			    $G_{yx}[i] \gets G_{x}[i]$\;
			    }
			\KwRet$G_{xy}, G_{yx}$
		\end{algorithm}
	\end{minipage}
\end{figure}

%

\vspace{.5em}
\noindent \textbf{Mutation.} The mutation is an operator commonly used in many genetic algorithms~\cite{Mitchell:1998}. It typically consists in making some bases of the DNA sequences of an individual to mutate into a different base (e.g., \texttt{A} $\xrightarrow{\makebox[1cm]{\scriptsize{mutation}}}$ \texttt{C}).

Previous studies showed that the distribution of bases within the digital DNA sequences of legitimate users is not uniform~\cite{dsaa2017,cresci2017social}. 
Quite intuitively, some actions tend to occur more often than others, such as the tweeting action. 
For this reason, in \textsc{GenBot} we design a mutation operator  that favors mutations from the \texttt{C} (replies) and \texttt{T} (retweets) bases to the \texttt{A} (tweets) base. Nonetheless, also mutations from \texttt{A} to \texttt{C} and \texttt{T} are possible, although with a lower probability. Our \texttt{Mutation}$(\cdot)$ function accepts two parameters: $P_i$, which is the full population at the $i$-th generation of the genetic algorithm; and $L$, which is the length of the digital DNA sequences. The output returned by the function is the mutated $P_i$ population.

\SetKwFunction{rand}{rand}
\SetKwData{Off}{Off}
\SetKw{Or}{or}
\begin{algorithm}[h]
	\scriptsize
	\NoCaptionOfAlgo
	\DontPrintSemicolon

	\caption{\texttt{Mutation}($\mathit{P_i  , L  }$)\label{alg:mut}}
	\For{ $ (G_j , v_j ) \in P_i $ }{
	  \For{ $ u_k \in G_j $ }{
	    \For{ $ l \in \{ 1 , \dots , L \} $ }{
	      $r \gets \rand()$\;
	      \If{ $ r < \textsf{\upshape MUT-PROB} $ }{
	        \eIf{ $u_k[l] = \mathtt{C}$ \Or $u_k[l] = \mathtt{T}$ }{
	          $u_k[l] = \mathtt{A} $}
	        {
	          \eIf{ $r < 0.5 $ }{
	            $u_k[l] = \mathtt{C} $
	          }{
	            $u_k[l] = \mathtt{T} $
	          }
	        }
	      }
	    }
	  }
	}
	\KwRet{ $P_i$ }
\end{algorithm}



\begin{figure*}[t]
	\centering
	\begin{minipage}[t]{0.5\textwidth}
		\centering
		\includegraphics[width=1\textwidth]{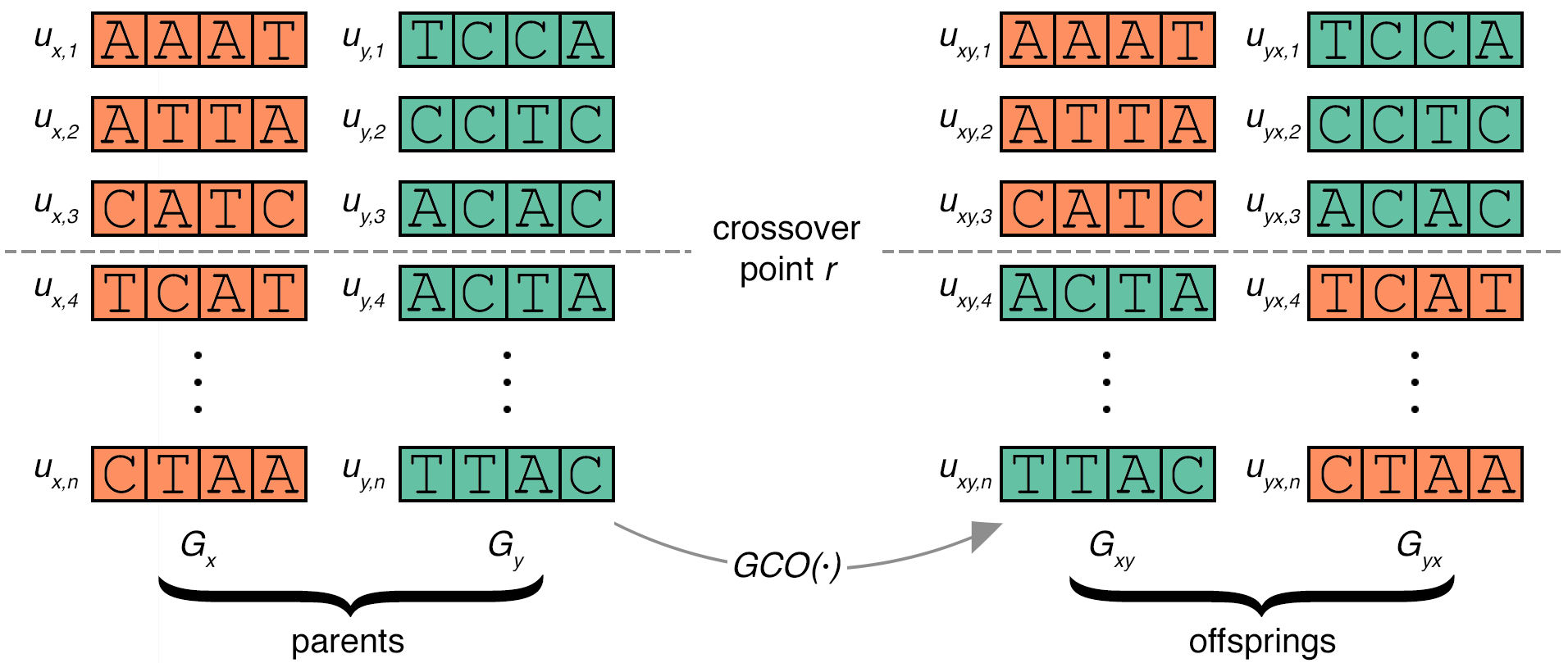}
		\caption{Group-level crossover: \texttt{\textmd{GCO}}$(\cdot)$ operator.\label{fig:group-crossover}}
	\end{minipage}
	\hspace{0.075\textwidth}%
	\begin{minipage}[t]{0.4\textwidth}
		\centering
		\includegraphics[width=1\textwidth]{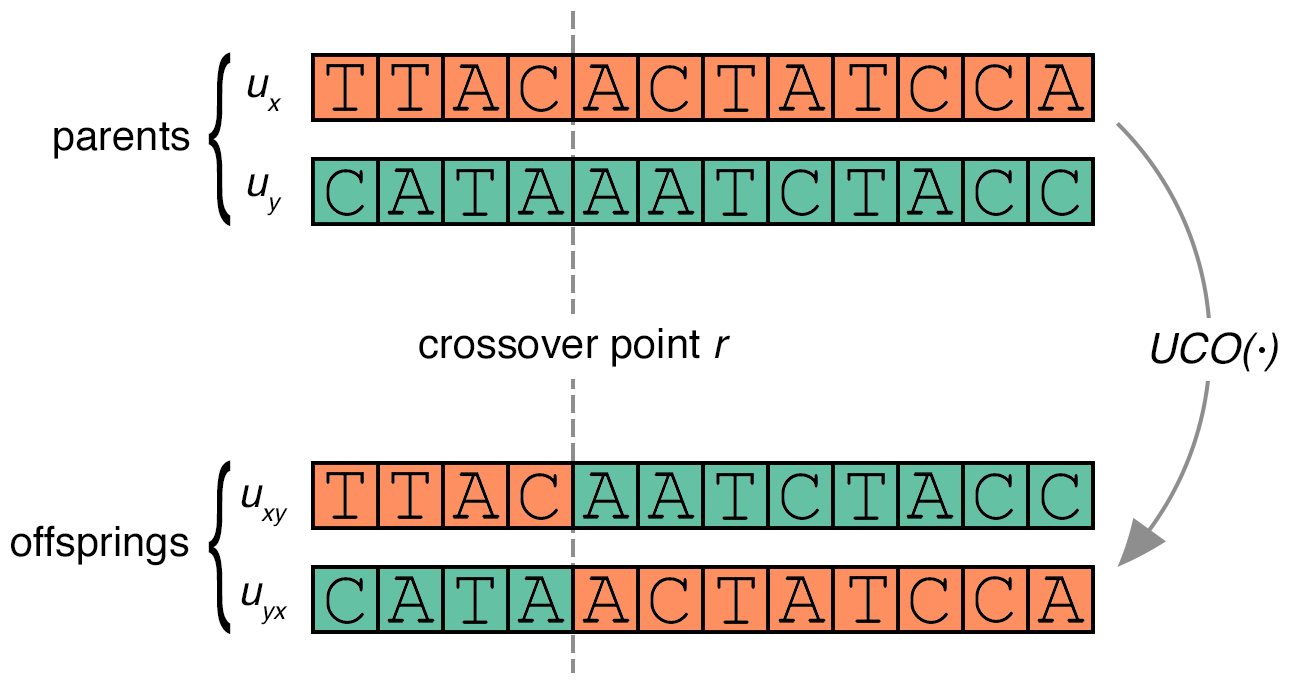}
		\caption{User-level crossover: \texttt{\textmd{UCO}}$(\cdot)$ operator.\label{fig:user-crossover}}
	\end{minipage}
\end{figure*}

\vspace{.5em}
\noindent \textbf{Crossover.} Traditionally, a crossover operator has two parent individuals as input, and generates two offsprings as the output~\cite{Mitchell:1998}. The offsprings are obtained via a recombination of the DNA sequences of the two parents. Remarkably, in this study an individual $G_j$ is a group of users $G_j = \{u_1, u_2, \ldots, u_M \}$ rather than a single user with its DNA sequence. Thus, by following the traditional crossover approach, we have two groups of users as input (the parents) and we obtain two groups of users as output (the offsprings). Recombinations at this level (i.e., at the group-level) can occur by mixing users between the two parent groups, rather than by mixing DNA sequences. Within \textsc{GenBot}, this crossover strategy is implemented with the function \texttt{GCO}$(\cdot)$ (\textit{\textbf{G}roup \textbf{C}ross\textbf{O}ver}). \texttt{GCO}$(\cdot)$ uses a one-point crossover technique. In detail, it randomly selects two groups $G_x, G_y \in P_i$ as the parents and a random crossover point $r$. Then, it generates the offspring $G_{xy}$ as the combination of the first parent up to point $r$ with the second parent from point $r$ onwards, and the offspring $G_{yx}$ viceversa. Figure~\ref{fig:group-crossover} gives an intuitive idea of how our group-level crossover operator works.




Since our individuals are composed of many users, 
we can complement the previous crossover strategy with additional fine-grained crossovers working at the user-level.
Specifically, we define two user-level crossover operators, referred to as the \textit{\textbf{U}ser \textbf{C}ross\textbf{O}ver} operator, defined in function \texttt{UCO}$(\cdot)$, and the \textit{\textbf{U}ser \textbf{R}everse \textbf{C}ross\textbf{O}ver} operator, defined in function \texttt{URCO}$(\cdot)$.
Function \texttt{UCO}$(\cdot)$ is a one-point crossover operator that acts at user-level. Similarly to the \texttt{GCO}$(\cdot)$ operator, two parents $u_x, u_y \in G_j$ and one crossover point $r$ are randomly picked. Next, two offsprings -- respectively $u_{xy}$ and $u_{yx}$ -- are generated as the combination of the first parent up to point $r$ with the second parent from point $r$ onwards, and viceversa. Figure~\ref{fig:user-crossover} gives an intuitive idea of how our \texttt{UCO}$(\cdot)$ user-level crossover operator works, in comparison with the group-level crossover.

Function \texttt{URCO}$(\cdot)$ differs from \texttt{UCO}$(\cdot)$ in the way the second parent $u_y$ is exploited. In fact, in \texttt{URCO}$(\cdot)$, the digital DNA sequence of $u_y$ is reversed before being recombined with that of the first parent $u_x$. This simple operation allows to create much more variability in the DNA sequences of the offsprings
and it demonstrates very effective in practice.

Contrarily to traditional genetic algorithms, we exploit the richness of the DNA-based behavioral representations, by designing a multi-level crossover strategy. At the user-level, the fine-grained \texttt{UCO}$(\cdot)$ and \texttt{URCO}$(\cdot)$ operators apply recombinations to the DNA sequences of single users. Furthermore, at the group-level, the coarse-grained \texttt{GCO}$(\cdot)$ operator shuffles users between different groups.

\begin{minipage}[t]{0.22\textwidth}
	\removelatexerror
	\vspace{0pt}
	\hspace{-0.4cm}
	\begin{algorithm}[H]
		\scriptsize
		\NoCaptionOfAlgo
		
		\caption{\texttt{UCO}($\mathit{ u_{x} , u_{y},r }$)\label{alg:inOnePt}}
		\For { $i \in \{ 1 , \dots , r \}$ }{
		    $u_{xy}[i] \gets u_{x}[i]$ \;
		    $u_{yx}[i] \gets u_{y}[i]$}
		\For{ $ i \in \{r+1 , \dots , |u_x| \} $  }{
		    $u_{xy}[i] \gets u_{y}[i]$ \;
		    $u_{yx}[i] \gets u_{x}[i]$}
		\KwRet$u_{xy}, u_{yx}$
	\end{algorithm}
\end{minipage}
\hspace{-0.2cm} 
\begin{minipage}[t]{0.22\textwidth}
	\removelatexerror
	\vspace{0pt}
	\begin{algorithm}[H]
		\scriptsize
		\NoCaptionOfAlgo
		
		\caption{\texttt{URCO}($\mathit{ u_{x} , u_{y},r }$)\label{alg:inCx}}
		\For{ $i \in \{1 , \dots , r \}$ }{
		    $u_{xy}[i] \gets u_{x}[i]$\;
		    $u_{yx}[i] \gets u_{y}[|u_y| - i]$}
		\For{ $i \in \{r+1 , \dots , |u_x| \} $}{
		    $u_{xy}[i] \gets u_{y}[|u_y| - i]$\;
		    $u_{yx}[i] \gets u_{x}[i]$}
		\KwRet$u_{xy}, u_{yx}$
	\end{algorithm}
\end{minipage}



\subsection{The \textsc{GenBot} algorithm}
\setcounter{algocf}{0}
\begin{algorithm}[t!]
	\scriptsize

	\SetKwData{Lcs}{$\mathtt{LCS}$}
	\SetKwData{Fit}{$\mathtt{Fit}$}
	\SetKwData{Mut}{$\mathsf{mut}$}
	\SetKw{Null}{null}
	\SetKwInOut{Input}{input}
	\SetKwInOut{Output}{output}
	\DontPrintSemicolon
	\caption{\textsc{GenBot}\label{alg:genbot}}
	
	\Input{target legitimate group $G_{\mathsf{legitimate}}$, initial population $P_0$}
	\Output{last generation of evolved spambots $P_{\mathsf{best}}$}
	\BlankLine
	$b \gets \Lcs(G_{\mathsf{legitimate}})$ \;
	$U \gets \mathtt{numOfUsers}(G_{\mathsf{legitimate}}) $ \;
	$T \gets \mathtt{numOfTweets}(G_{\mathsf{legitimate}}) $ \;
	$P_{\mathsf{best}} \gets \Null $ \;
	\For{ $ G_i \in P_0 $ }{
	    $v_i \gets \Fit(G_i , b )$ \tcp*[f]{initial fitness score} \;
	}
	\For{ $ i \in \{ 1 , \dots , \textsf{\upshape MAX-GEN} \} $ }{
	  $\Mut = \mathtt{Mutation}( P_{i-1} , T )$ \tcp*[f]{apply mutations} \;
	  \For{ $ m_j \in \Mut $ }{
	    $ mv_j \gets \Fit ( m_j , b ) $ \;
	    \If{ $mv_j < v_j$ }{
	      $(G_j,v_j) \gets (m_j,mv_j)$ \;
	     }
	  }
	  \For{ $ (G_k,v_k) \in P_i $ }{
	    $x \gets \rand(1,\textsf{POP-SIZE})$\;
	    $y \gets \rand(1,\textsf{POP-SIZE})$ \;
	    $G_{xy} \gets \mathtt{GCO}( G_{x} , G_{y} , U )$ \tcp*[f]{apply group crossovers} \;
	    \For(\tcp*[f]{apply user reverse crossovers}){ $j \in \{ 1 , \dots , \textsf{\upshape NUM-URCO} \}$}
	    {
	     $u_x \gets \rand(1,U)$\;
	     $u_y \gets \rand(1,U)$\;
	     $(u_{xy},u_{yx}) \gets \mathtt{URCO}( G_{j}[u_x] , G_{j}[u_y] , T )$  \;
	     $G_{xy}[u_x] \gets u_{xy}$ \;
	     $G_{xy}[u_y] \gets u_{yx}$ \;
		}
		$ mv_k \gets \Fit ( G_{xy} , b )$ \;
	 	\If{ $mv_k < v_k $ } {
		 $ (G_k,v_k) \gets (G_{xy} , mv_k) $ \;
		 }
	    }
	  \For{ $ (G_k , v_k) \in  P_i $ }{
	    \For(\tcp*[f]{apply user crossovers}){ $j \in \{ 1 , \dots , \textsf{\upshape NUM-UCO} \}  $  }
	    {
	     $u_x \gets \rand(1,U)$\;
	     $u_y \gets \rand(1,U)$\;
	    $(u_{xy},u_{yx}) \gets \mathtt{UCO}( G_{j}[u_x] , G_{j}[u_y] , T )$  \;
	     $G_{xy}[u_x] \gets u_{xy}$ \;
	     $G_{xy}[u_y] \gets u_{yx}$ \;
	    }
	    $mv_k \gets \Fit ( G_{xy} , b ) $ \;
	    \If{ $mv_k < v_k $ }
	    {
	    $ (G_k , v_k) \gets ( G_{xy} , mv_k ) $ \;
	    }
	    }
	  $P_{i+1} \gets P_{i}$ \;
	  $P_{\mathsf{best}} \gets P_{i}$ \tcp*[f]{update evolved spambots} \;
	}
	\KwRet{$P_{\mathsf{best}}$}
\end{algorithm}
We implemented \textsc{GenBot} by following the steps of the (1+1)-evolutionary algorithm scheme~\cite{Droste:2002}. In its simplest definition, a (1+1)-EA is a \textit{randomized hill climbing technique}~\cite{Wattenberg:CSD-94-834} that relies on mutations only. In the \textsc{GenBot} algorithm, the simple (1+1)-EA scheme is extended by the adoption of a multi-level crossover strategy (i.e., 1 group-level and 2 user-level crossovers), as previously defined.

The core of the \textsc{GenBot} algorithm is represented in Algorithm~1 by the \textit{for loop} at lines 7--37. Each iteration of the loop applies the mutation and crossover operators  to obtain a new generation of spambots. In detail, the evolutionary steps in \textsc{GenBot} begin by mutating the current population and continue by substituting the individuals that improve as a consequence of the mutations. Then, the group-level (coarse-grained) crossover \texttt{GCO}$(\cdot)$ is applied. Finally, also the 2 user-level (fine-grained) crossovers are applied. Specifically, at first \texttt{URCO}$(\cdot)$ is applied and the obtained offsprings are evaluated. Only those offsprings that improved the fitness score are retained. Subsequently, \texttt{UCO}$(\cdot)$ is applied and the offsprings are evaluated one last time, thus obtaining a new population that becomes the starting point of the next iteration of the algorithm.

Notably, traditional evolutionary heuristics based on genetic algorithms perform only one update of the population at each iteration of the algorithm. In \textsc{GenBot} however, each generation is the combination of three intermediate generations, respectively obtained via (i) mutations, (ii) a combination of group-level and reverse user-level crossovers, and (iii) user-level crossovers.




\section{Experiments, setup and evaluation}
\label{sec:exp}

\subsection{Dataset}
\label{sec:twitter-dataset}
The dataset for this study is composed of the timelines of 3,474 legitimate Twitter accounts.

In order to build this dataset of certified human-operated accounts, random Twitter users were contacted by mentioning them in tweets. Then, contacted users were asked simple questions in natural language. Possible answers to such questions were collected by means of a Twitter crawler\footnote{\url{https://developer.twitter.com/en/docs}}. Upon manually verifying the answers, all 3,474 accounts that answered were certified as legitimate ones.
Notably, this dataset has already been used in recent works~\cite{cresci2016dna,dsaa2017,Cresci2017,cresci2017social} and it is considered an important resource in the field of spambot and automation detection.

\subsection{Experimental setup}
\label{sec:experiments-setup}
The \textsc{GenBot} algorithm is implemented in C++ and the code is publicly available for scientific purposes (link in the introduction).
For an efficient, linear-time computation of the LCS curves, we rely on an adapted version of the GLCR toolkit\footnote{\url{https://www.uni-ulm.de/in/theo/research/seqana.html}} implementing the algorithms in~\cite{arnold2011}.
All the experiments ran on a machine with an Intel Xeon E7-4830v4, with a 64-bits architecture at 2 GHz, 112 cores and 500 GB of RAM.
As the reference with which to compare our results, we consider the last (most recent) 2,000 actions performed by the legitimate accounts in our dataset. 

We run \textsc{GenBot} with a population of 30 individuals per run (\textsf{POP-SIZE}) and a generation limit of 20,000 epochs as a stopping criterion (\textsf{MAX-GEN}).
The initial population $P_0$ is composed of 30 identical individuals. Each individual represents a group of accounts and the starting point for all the individuals is a DNA sequence of length 2,000 (same DNA length of the legitimate accounts), whose first 1,000 positions are filled with the DNA base \verb|A|, followed by 500 positions filled with the base \verb|C| and the last 500 positions filled with the base \verb|T|.
Regarding the mutation operator, the probability to mutate each action is set equal to $0.0002$ (\textsf{MUT-PROB}).
For each generation simulated by \texttt{GenBot}, a total of 30 offsprings are generated via group-level crossovers. Concerning the user-level crossovers, for each individual 2 offsprings are generated via the \texttt{URCO}$(\cdot)$ operator (\textsf{NUM-URCO}), while 12 offsprings are generated via the \texttt{UCO}$(\cdot)$ operator (\textsf{NUM-UCO}). 
Each experiment is repeated 5 times and each run of the algorithm starts with a different random seed. Results are averaged across the 5 runs.

\subsection{Experimental evaluation}
\label{sec:results}
In the next section, we provide results for our experiments.
In each experiment, the quality of the solutions generated by the the \texttt{GenBot} algorithm is evaluated by comparing the LCS curve of the last generation of spambots with that of the reference group. The LCS curve of the last generation of spambots is computed by applying the point-to-point average of all the LCS curves of the spambots groups constituting the last generation.
We give both a qualitative and a quantitative evaluation of the results, as follows: (i) we provide a graphical comparison of LCS curves, giving a direct and intuitive insight into the quality of our results; (ii) we compute and compare the AUC of the LCS curves with Equation~(\ref{eq:sequence-def}); (iii) we measure the distance between the LCS curves by means of the $D_{KL}$ defined in Equation~(\ref{eq:kl-distance}).



\begin{figure*}[t]
	\begin{minipage}[t]{0.235\textwidth}
		\centering
		\includegraphics[width=1\textwidth]{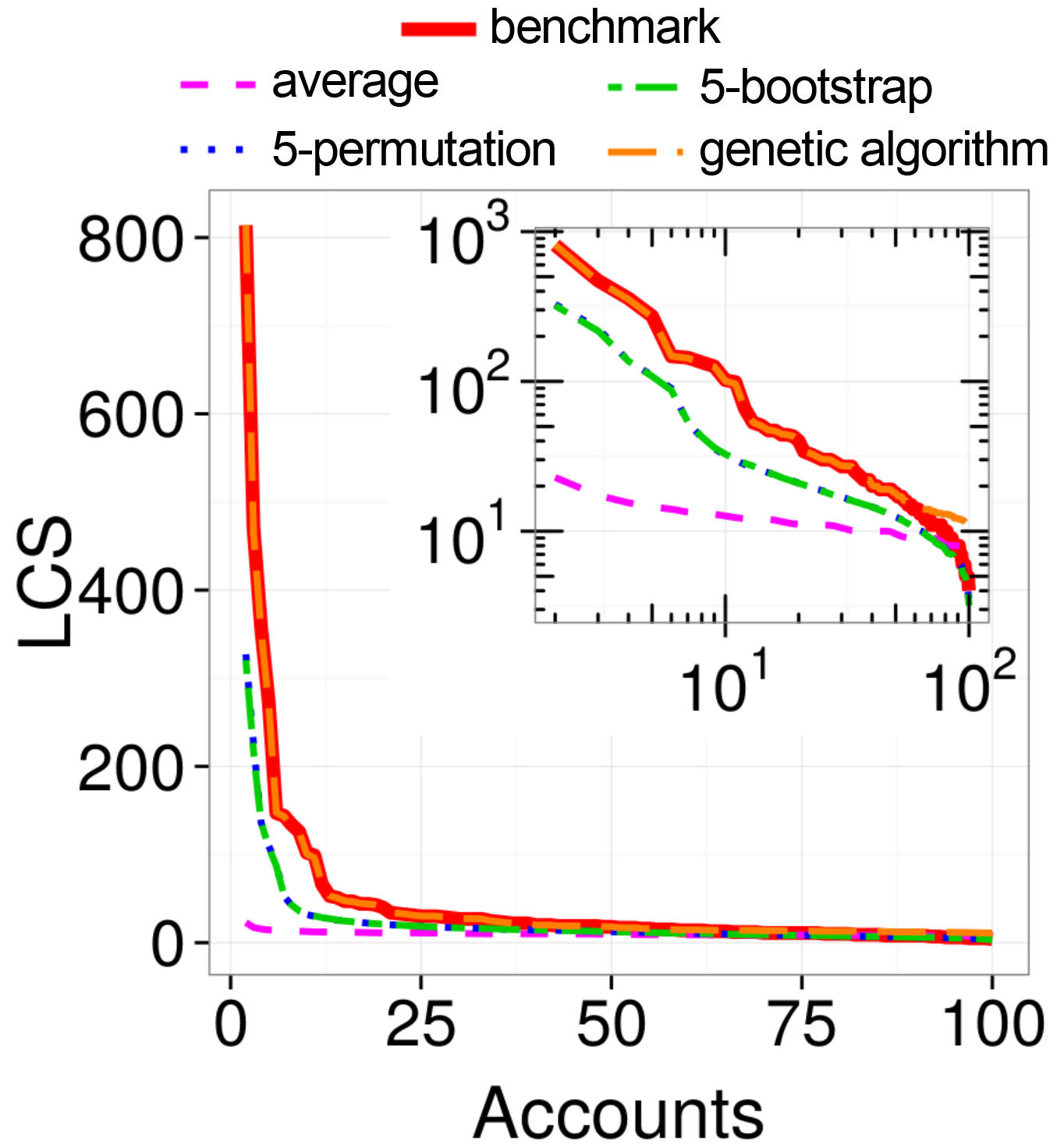}
		\caption{Qualitative analysis of evolved spambots and comparison with previous techniques.\label{fig:main}}
	\end{minipage}
	\hspace{0.04\textwidth}
	\begin{minipage}[t]{0.235\textwidth}
		\centering
		\includegraphics[width=1\textwidth]{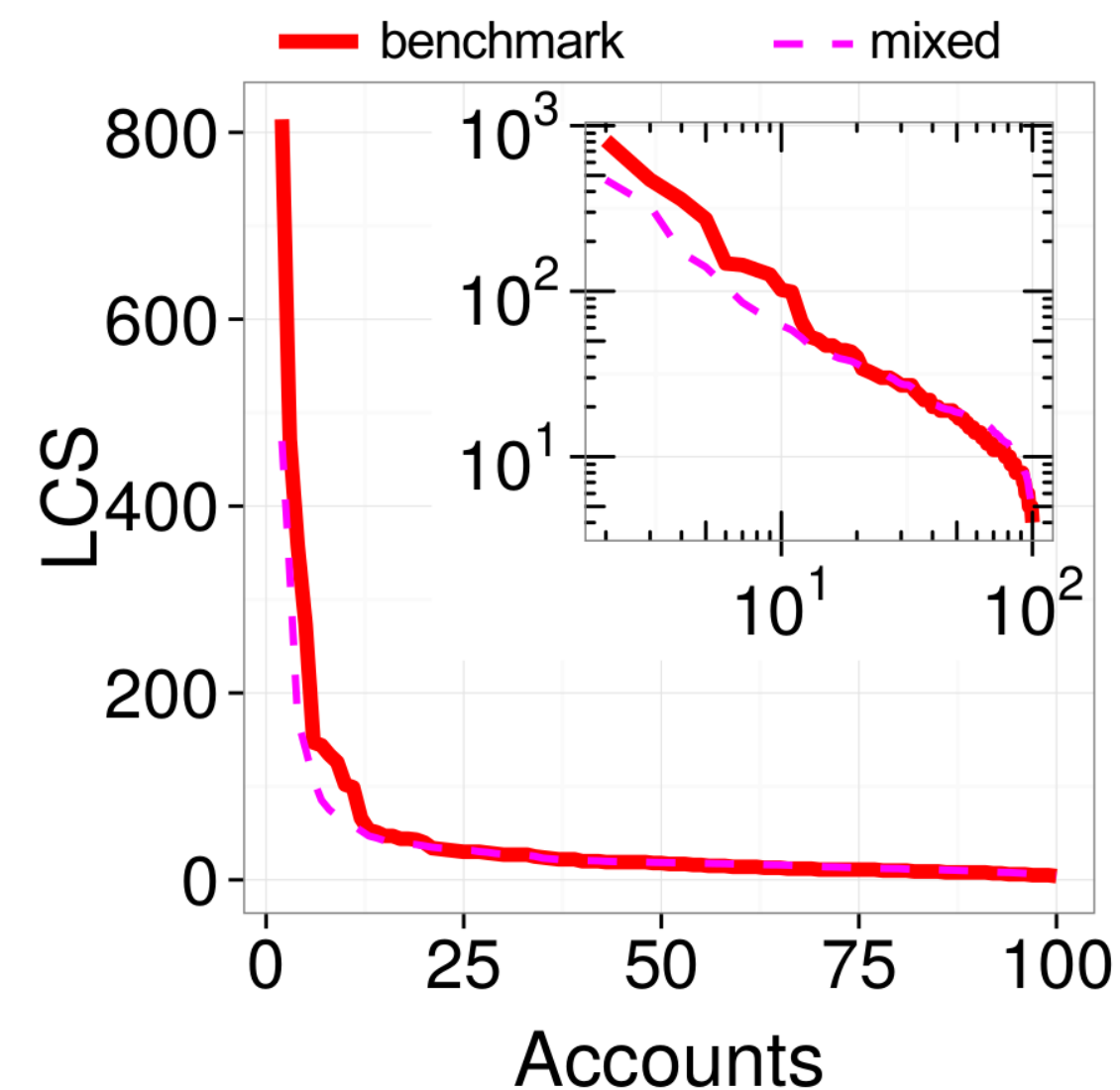}
		\caption{Comparison of the LCS curve of legitimate users with that of a mixed group composed of evolved spambots and legitimate users.\label{fig:runMix}}
	\end{minipage}
	\hspace{0.04\textwidth}
	\begin{minipage}[t]{0.43\textwidth}
		\centering
		\includegraphics[width=1\textwidth]{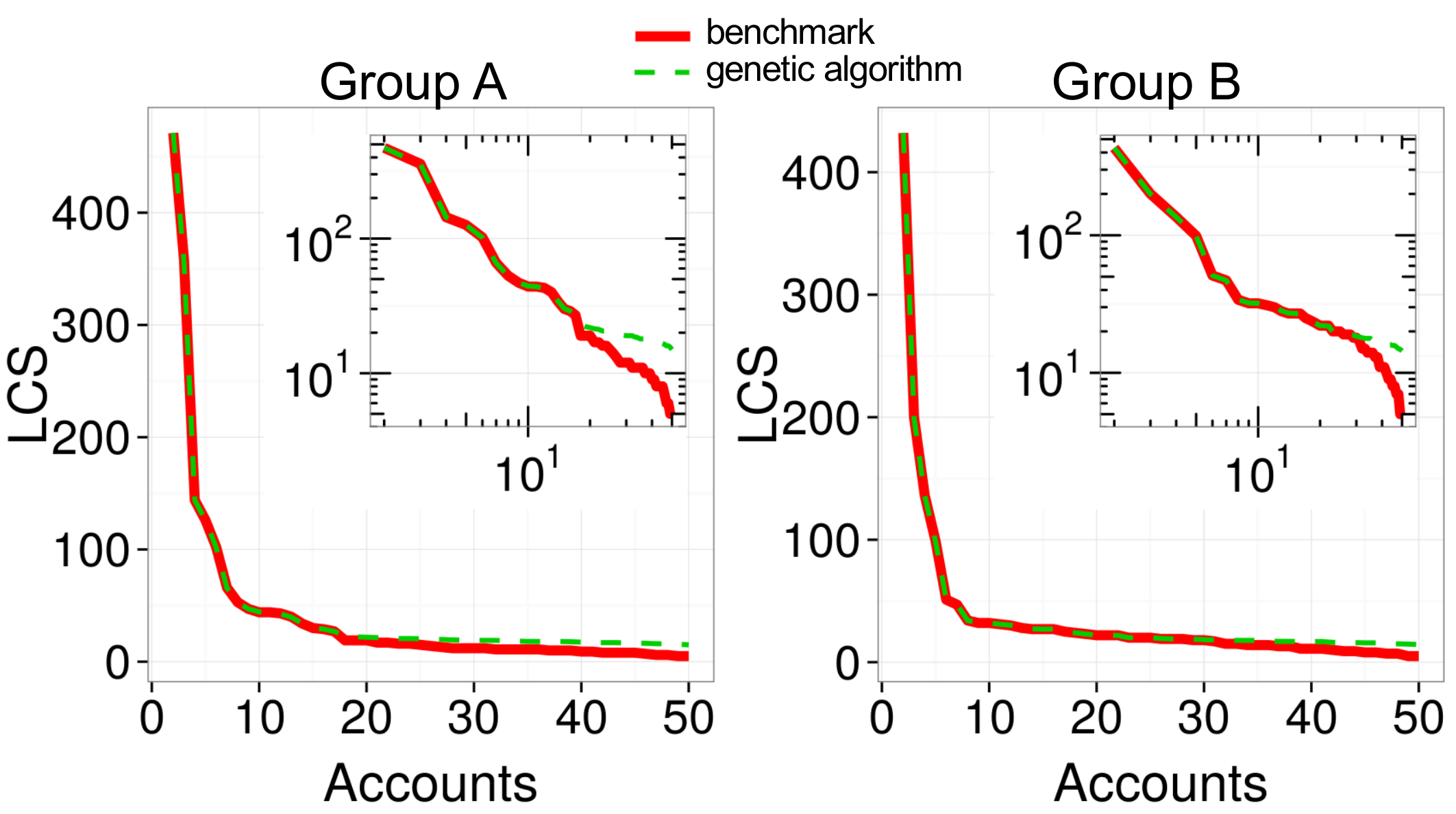}
		\caption{Qualitative comparison of evolved spambots against 2 different groups of legitimate accounts. Despite the different shape of the LCS curves of \emph{Group A} and \emph{Group B}, the corresponding evolved spambots closely match their behavior.\label{fig:run50B}\label{fig:run50A}}
	\end{minipage}
\end{figure*}

\section{Results}
\label{sec:results}
The last generation of spambots generated by \textsc{GenBot} (i.e., the one under evaluation) is referred to as \textit{evolved spambots}.

\subsection{Behavioral analysis of evolved spambots}
\label{sec:res-behavior}
\begin{figure}[h]
  \centering
  \includegraphics[width=0.9\columnwidth]{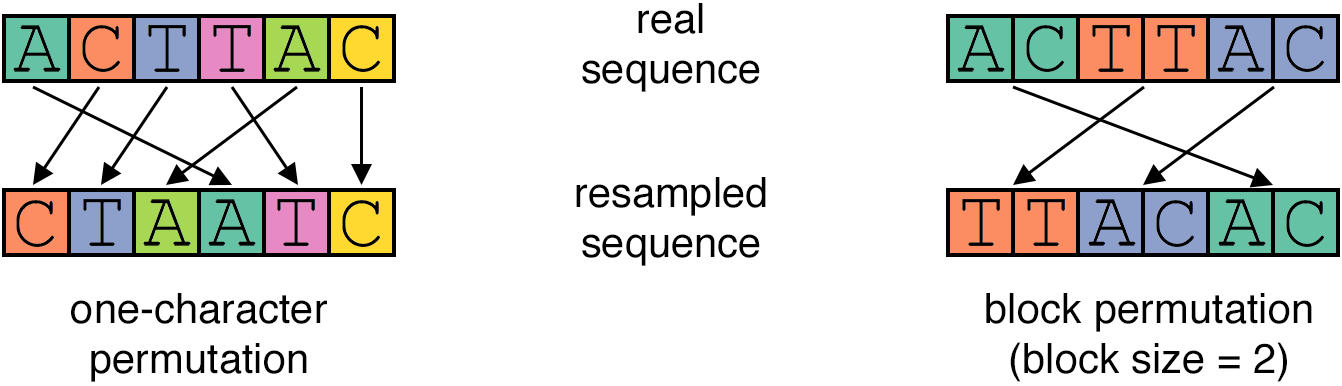}
  \caption{Application of block resampling to a digital DNA sequence, and comparison with one-character resampling.
  \label{fig:block-perm-example}}
\end{figure}

Here, we evaluate the extent to which the evolved spambots generated by \textsc{GenBot} are capable of emulating the behavior of legitimate users.
The first -- and to the best of our knowledge, unique -- scientific attempt to solve this task was documented in~\cite{dsaa2017}, where the authors employed a set of resampling techniques to generate new behavioral fingerprints, as similar as possible to those of legitimate users. In order to provide a comparison between \textsc{GenBot} and~\cite{dsaa2017}, we applied the best performing techniques described in~\cite{dsaa2017} to our dataset. Specifically, 3 types of DNA resampling techniques were used: (i) a statistical resampling of the digital DNA sequences of legitimate users based on the average characteristics of those users (labeled \textit{average}); (ii) a block permutation with block size $= 5$ (labeled \textit{5-permutation}); and (iii) a block bootstrap with block size $= 5$ (labeled \textit{5-bootstrap}). For the sake of clarity, 
Figure~\ref{fig:block-perm-example} pictorially shows the way to perform a block resampling on a digital DNA sequence~\cite{dsaa2017}.

Qualitative results of this experiment are shown in Figure~\ref{fig:main}. The LCS curve of the group of legitimate accounts is labeled \textit{benchmark}. 
As shown, the LCS curve of the spambots generated with \textsc{GenBot} is almost completely overlapping the benchmark. The only exception is in the tail of the LCS curve and it is visible in the log-log inset of Figure~\ref{fig:main}. The comparison between previous techniques and \textsc{GenBot} is clearly in favor of the latter. In fact, all previous techniques greatly struggle to fit the head of the LCS curve, while instead they perform better in the tail. To this regard, our approach and those described in~\cite{dsaa2017} seem to be complementary, with our \textsc{GenBot} algorithm capable of fitting all the LCS curve except for the tail, which instead is well fit by the resampling approaches of~\cite{dsaa2017}. 
One further improvement could be the adoption of block resampling strategies in \textsc{GenBot}, adding them to the mutations and crossovers that we already employ.

A quantitative evaluation of the quality of evolved spambots is obtained by comparing the AUC of the LCS curves of the spambots with that of the legitimate accounts. 
As shown in Table~\ref{tab:auc}, the spambots obtained with \textsc{GenBot} were able to reproduce the LCS of legitimate users with a percentage error of only 2.98\% in excess. Instead, all other techniques managed to reproduce only about half of the behavioral similarities expected within a group of legitimate accounts.



\begin{table*}[t]
	\scriptsize
	\centering
	\begin{minipage}[t]{0.235\textwidth}
          \centering
          \setlength{\tabcolsep}{4pt}
		\begin{tabular}{lrr}
			\toprule
			\textbf{technique} & \textbf{AUC} & \textbf{\% error} \\
			\midrule
			benchmark		& 3961.0	& -- \\
			\midrule
			average			& 971.9	& $-$75.46\% \\
			5-bootstrap		& 2001.7	& $-$49.46\% \\
			5-permutation		& 2019.8	& $-$49.01\% \\
			genetic algorithm	& 4079.0	& $+$2.98\% \\
			\bottomrule
		\end{tabular}
		\caption{Quantitative analysis: comparison with previous techniques.\label{tab:auc}}
	\end{minipage}
	\hspace{0.04\textwidth}
	\begin{minipage}[t]{0.275\textwidth}
          \centering
          \setlength{\tabcolsep}{4pt}
		\begin{tabular}{lrr}
			\toprule
			\textbf{group} & \textbf{AUC} & \textbf{\% error} \\
			\midrule
			benchmark (legitimate)	& 3961.0		& -- \\
			mixed (bot + legitimate)	& 3090.2		& $-$21.98\% \\
			\bottomrule 
		\end{tabular}
		\caption{AUC comparison (group of legitimate users vs mixed groups: evolved spambots and legitimate users).\label{tab:aucHy}}
	\end{minipage}
	\hspace{0.04\textwidth}
	\begin{minipage}[t]{0.39\textwidth}
          \centering
                    \setlength{\tabcolsep}{4pt}
		\begin{tabular}{lrrcrr}
			\toprule
			& \multicolumn{2}{c}{\textit{Group A}} && \multicolumn{2}{c}{\textit{Group B}} \\
			\cmidrule{2-3} \cmidrule{5-6}
			\textbf{technique} & \textbf{AUC} & \textbf{\% error} && \textbf{AUC} & \textbf{\% error} \\
			\midrule
			benchmark 		& 1797.0	& --			&& 1521.50	& -- \\ 
			genetic algorithm	& 2025.6	& $+$12.72\%	&& 1632.20	& $+$07.28\% \\ 
			\bottomrule
		\end{tabular}
		\caption{Quantitative comparison of evolved spambots against 2 different groups of legitimate accounts.\label{tab:auc50}}
	\end{minipage}
\end{table*}

\begin{table*}[t]
	\scriptsize
	\centering
	\begin{minipage}[t]{0.65\textwidth}
          \centering
          \setlength{\tabcolsep}{4pt}
		\begin{tabular}{ll@{\phantom{M}}rrrrcr}
			\toprule
			&& \multicolumn{6}{c}{\textbf{evaluation metrics}} \\
			\cmidrule{3-8}
			\textbf{technique} & \textbf{accounts} & \textit{Precision} & \textit{Recall} & \textit{Specificity} & \textit{Accuracy} & \textit{F1} & \textit{MCC}\\
			\midrule
				Cresci \textit{et al.}~\cite{cresci2017social}		& non-evolved spambots				& 1.000	& 0.858	& 1.000	& 0.929	& 0.923	& 0.867 \\
				Miller \textit{et al.}~\cite{miller2014} 				& non-evolved spambots				& 0.555	& 0.358	& 0.698	& 0.526	& 0.435	& 0.059 \\
			\midrule
				Cresci \textit{et al.}~\cite{cresci2017social}		& evolved spambots (\textsc{GenBot})	& 0.512	& 0.210	& 0.800	& 0.505	& 0.298	& 0.012 \\
				Miller \textit{et al.}~\cite{miller2014} 				& evolved spambots (\textsc{GenBot})	& 0.720	& 0.360	& 0.860	& 0.610	& 0.480	& 0.254 \\
			\bottomrule
		\end{tabular}
		\caption{Performances of 2 state-of-the-art spam and bot detection techniques towards the detection of non-evolved spambots and evolved spambots generated with \textsc{\textmd{GenBot}}. The evolved spambots largely go undetected.\label{tab:detection}}
	\end{minipage}
	\hspace{0.04\textwidth}
	\begin{minipage}[t]{0.3\textwidth}
		\centering
		\begin{tabular}{lcrr}
			\toprule
			&& \multicolumn{2}{c}{$D_{KL}$} \\
			\cmidrule{3-4}
			\textbf{target group} && \textbf{mean} & \textbf{std} \\
			\midrule
			legitimate full 			&& 32.64	& 2.97 \\
			legitimate \textit{Group A}	&& 73.83	& 11.44 \\
			legitimate \textit{Group B} 	&& 34.16	& 2.33 \\
			\bottomrule
		\end{tabular}
		\caption{Variability of our results across 5 runs of our algorithm for different experiments.\label{tab:variability}}
	\end{minipage}
\end{table*}

\subsection{Detection of evolved spambots}
\label{sec:res-detection}
Going further, we are now interested in evaluating whether the evolved spambots are able to avoid detection by state-of-the-art techniques~\cite{cresci2017social}.
We design this experiment by replicating the working conditions of most spambot detection systems -- that is, we focus on the analysis of an unknown group of users that contains both spambots and legitimate users. 


In detail, we start by mixing together part of our evolved spambots with part of the legitimate users. Then, we compare the LCS curve of the mixed group with that of the legitimate users only. Figure~\ref{fig:runMix} shows a qualitative result of this comparison. As shown, the LCS curve of the mixed group lays very close to that of the legitimate users. In turn, this means that also the mixed group of evolved spambots and legitimate users still behaves like a group solely composed of legitimate users. Table~\ref{tab:aucHy} presents the comparison in terms of AUC values, which quantitatively confirm the result, although showing a larger error than that reported in the previous experiment of Table~\ref{tab:auc}.

Finally, we apply 2 state-of-the-art spam and bot detection techniques~\cite{miller2014,cresci2017social} to the mixed group, and we assess their performance in detecting the evolved spambots. We compare these results with those measured while applying the techniques in~\cite{miller2014,cresci2017social} to a group of non-evolved spambots. Results are reported in Table~\ref{tab:detection} and show that the evolved spambots generated by \textsc{GenBot} largely evade detection (mean $F1 \simeq 0.260$). In addition to the techniques tested in Table~\ref{tab:detection}, we also applied the system in~\cite{ahmed2013} to our evolved spambots. Similarly to~\cite{miller2014,cresci2017social}, also~\cite{ahmed2013} proves incapable of accurately detecting the evolved bots with $Accuracy = 0.495$ and $MCC = -0.071$. This result is in contrast with previous work~\cite{cresci2016dna,cresci2017social} where the detection rate for non-evolved spambots was $F1 = 0.923$.


\subsection{Generalizability}
\label{sec:res-general}
In this section, we evaluate the generalizability of the \textsc{GenBot} algorithm and of the previously shown results.
We change the group of legitimate accounts and we assess whether the evolved spambots generated by \textsc{GenBot} are still similar to the legitimate accounts.
We first randomly split the original group of legitimate accounts into 2 disjunct subgroups (labeled \emph{Group A} and \emph{Group B}), each subgroup counting about 50\% of the accounts of the original group.
These subgroups are the 2 new references for \textsc{GenBot} to generate evolved spambots. Then, we compare the behavioral similarity between the evolved spambots and the subgroup (either \emph{Group A} or \emph{Group B}) used to generate them. Figure~\ref{fig:run50A} shows the results of a qualitative comparison. Despite the different shape of the LCS curves of \emph{Group A} and \emph{Group B} of legitimate accounts (solid red line), the corresponding evolved spambots closely match their behavior. This testifies that even by changing the characteristics of the accounts to mimic, \textsc{GenBot} is capable of generating spambots that behave in a similar way with respect to the legitimate ones. Moreover, this also implies that \textsc{GenBot} is capable of generating spambots featuring different characteristics.
Interestingly, Figure~\ref{fig:run50A} also shows a rather poor performance in fitting the tail of the LCS curve, similarly to what we already saw in Figure~\ref{fig:main}.
Finally, Table~\ref{tab:auc50} reports quantitative results for the 2 subgroups of legitimate accounts, showing a low percentage error for both groups.

%


The experiments
were executed 5 times, with random seeds to assess variability. 
In particular, we measure the mean and the standard deviation (std) of the distance between the LCS curve of the evolved spambots generated by \textsc{GenBot} and that of legitimate accounts, across 5 runs of the algorithm. The distance between LCS curves is measured by means of the $D_{KL}$ distance defined in Equation~(\ref{eq:kl-distance}). As shown in Table~\ref{tab:variability}, the standard deviation of $D_{KL}$ is rather low in every experiment, and significantly lower than the mean, which suggests low variability in the results.

\subsection{Improving current detection techniques}
\label{sec:res-entropy}

\begin{figure}[t]
\centering
\includegraphics[width=0.35\textwidth]{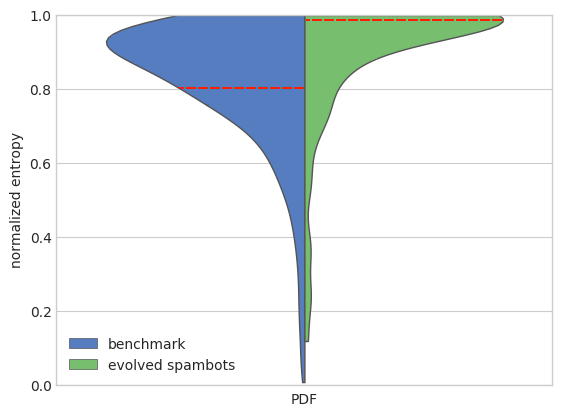}
\caption{Beanplot showing the PDF of the normalized Shannon entropy of DNA sequences related to evolved spambots and legitimate accounts. DNA sequences of the evolved spambots feature a suspiciously high entropy.\label{fig:entropy}} 
\end{figure}

As demonstrated above, the \textsc{GenBot}-generated spambots are able to avoid actual detection techniques. We can see that
collectively, the evolved spambots do not leave traces of their automated nature, since they accurately reproduce the LCS curve of legitimate accounts. However, interesting insights can be gained by studying the single digital DNA sequences of every spambot. Indeed, a close inspection reveals that the digital DNA sequences of the spambots generated by \textsc{GenBot} have very few repetitions and almost no regularities at all\footnote{Here we are interested in repetitions \textit{within} sequences rather than \textit{across} sequences, as it was for the case of studying the LCS.}. In turn, such DNA sequences represent very erratic and heterogeneous behaviors. This finding is in contrast with the known characteristics of legitimate accounts~\cite{dsaa2017,cresci2017social} that tend to favor certain actions in their behaviors. As a consequence, digital DNA sequences of legitimate accounts have a prevalence for certain DNA bases (e.g., the \texttt{A} base for tweets).
We formalize this intuition by relying on the notion of normalized Shannon entropy ($H_{norm}$). In particular, we compute the entropy of each digital DNA sequence, for all the evolved spambots and all the legitimate accounts. Figure~\ref{fig:entropy} shows a beanplot of the empirical probability density function (PDF) of the normalized entropy, comparing measurements for evolved spambots with those of legitimate accounts. As expected, the spambots generated by \textsc{GenBot} have mean $H_{norm} \simeq 1$, whereas for legitimate accounts mean $H_{norm} = 0.8$ (dashed red lines in Figure~\ref{fig:entropy}).
A straightforward consequence of this observation is the possibility to extend current detection techniques based on the accounts' behavior, such as~\cite{cresci2016dna,cresci2017social}, by also considering information related to the repetitions and regularities within sequences, thus making current systems more robust against possible future spambot evolutions. Although limited in scope, this experiment nonetheless testifies the usefulness -- and one among many possible applications -- of the proposed analytical framework for simulating spambot evolutions.



\section{Discussion}
\label{sec:discussion}
Our results  demonstrated the possibility to combine the behavioral representation of digital DNA with the computational framework of genetic algorithms, in order to create evolved spambots capable of escaping current state-of-the-art detection techniques, based on the accounts' behavior~\cite{cresci2017social} and on the content of posts~\cite{miller2014,ahmed2013}. 

In particular, we designed an analytical framework for simulating spambot evolutions, thus answering to the first of our research questions (\textbf{RQ1}). In such framework, social spambots behavior is modeled via digital DNA. Then, DNA sequences are fed to the novel genetic algorithm (\textsc{GenBot}), designed to simulate spambot evolutions. After thousands of subsequent iterations, the output of \textsc{GenBot} is a novel generation of \textit{evolved spambots}, described by their digital DNA. 
Notably, in this work we grounded the framework for simulating possible spambot evolutions on (i) genetic algorithms and (ii) the recent advances in digital DNA behavioral modeling, since they currently represent its key enabling factors. However, it is likely that in the near future the same methodological approach for studying spambot evolution could leverage different techniques and methodologies, thus widening its applicability.

Aiming to answer \textbf{RQ2}, we then evaluated the extent to which the evolved spambots are capable of going undetected by state-of-the-art techniques. Since our experiments grounded on modeling the actions in the timelines of the accounts under investigation, it was natural to consider detection techniques based on accounts behavior and posted contents. Thus, with regards to the technique in~\cite{cresci2017social}, we showed that (i) the behavioral fingerprint of the spambots generated by \textsc{GenBot} is similar to that of legitimate users, (ii) a group containing both our evolved spambots and legitimate users is almost indistinguishable from a group solely composed of legitimate users, and (iii) the \textit{social fingerprinting} detection technique largely fails in detecting the evolved spambots. 
Moreover, 2 other recent detection techniques, based on the content of posts~\cite{miller2014,ahmed2013}, also fail in detecting the spambots generated by \textsc{GenBot}. These results raise concerns towards the vulnerabilities of current state-of-the-art techniques.

Finally, we studied the characteristics of the evolved spambots, with the goal of answering to \textbf{RQ3} -- that is, looking for ways to improve (at least a subset of) current detection techniques. Specifically, we investigated whether the evolved spambots still had some peculiar characteristics that would make them detectable. We noticed that, although the group of evolved spambots behaves like a group of legitimate users, the digital DNA of the spambots is more \textit{entropic} than that of legitimate users. Thus, we argue that it would be possible and fruitful to extend current detection techniques by also considering the amount of entropy within digital DNA sequences. This last finding thus represents a useful suggestion for improving current spambot detection techniques. 

This study -- the very first of its kind -- moves in the direction of a \textit{proactive} spambot detection. For the first time since the advent of OSNs, we have the chance to proactively study spambot evolutions and to design more robust detection techniques, possibly capable of withstanding the next evolutions of social spambots.
Although unlikely to completely defeat spambots and other malicious accounts, the application of the proposed proactive approach would nonetheless bring groundbreaking benefits.
The capability to foresee possible spambot evolutions would not only allow to test the detection rate of state-of-the-art techniques (including techniques based, e.g., on the exploration of the social graph of the accounts, or on their profiles and posting aptitudes), but also and above all, to \textit{a priori} adapt them and even to design new detection techniques.
As a consequence of the additional design and experimentation allowed by the proactive approach, many spambot evolutions will be detected from \textit{day 0}. Overall, spambots will see their chances to harm severely restricted, with clear and immediate benefits for our online environments, and ultimately, for our societies (e.g., less fake news and biased propaganda). Notably, for those few spambot evolutions still not foreseen by this proactive approach, we will still be able to fall back to the traditional reactive approach, at no additional cost.

\section{Ethical considerations}
\label{sec:ethics}
With the rise of AI, our daily lives are increasingly influenced by decisions taken on our behalf by automated systems.
Algorithmic filtering (which leads to filter bubbles, echo chambers, and eventually polarization), algorithmic bias and current limits in explainability of predictive models already raise serious ethical concerns on the development and adoption of AI solutions.

Within this context, algorithmic approaches to the characterization, development, and detection of social bots make no exception~\cite{thieltges2016devil,de2017social}.
For instance, one might naively think that all endeavors devoted to the development of social bots are to be blamed.
Remarkably, however, not all social bots are nefarious by nature. Indeed, as highlighted in~\cite{ferrara2016}, bots can be programmed to automatically post information about news and academic papers~\cite{Hausten2016,Lokot2016}, and even to provide help during emergencies~\cite{savage2016botivist,avvenuti2017hybrid}. Undeniably, the provision of useful services by benign bots will make them become an established presence on social platforms~\cite{Monsted2017EvidenceOC}. Meanwhile, other researchers investigated the development and the behaviors of malicious bots~\cite{Lee2011seven,aiello2012people,cresci2019capability,websci17}, in an effort to better understand their evolution, impact and interactions with benign users.
The so-recognized existence of benign versus malicious bots sparked heated debates about the rights of automated accounts. For instance, both researchers and everyday social media users wondered the extent to which social bots should be considered equal to humans, as far as censorship\footnote{\url{https://www.nytimes.com/2018/09/05/technology/lawmakers-facebook-twitter-foreign-influence-hearing.html}}\textsuperscript{,}\footnote{\url{https://www.theguardian.com/technology/2018/oct/16/facebook-political-activism-pages-inauthentic-behavior-censorship}} and suppression of free speech~\cite{Mervis2014} are concerned.


 
While advancing the state of the art in the fascinating field of AI, we are aware that the successful implementation of new technologies will pose greater challenges to discriminate between human and automated behaviors. Now more than ever, spambot evolution and their subsequent detection meet ethical considerations. The framework proposed in this paper should not be seen merely as a technical exercise. In fact, our evolved spambots have not been conceived to support botnet developers (a criticism that could be very well posed to all the above-cited research). Instead, we remark that one of the main goals of this paper is to proactively sharpen detection techniques to cope with future evolutions of spambots, as typically done in the well-recognized field of adversarial learning.

Lastly,~\cite{Ferrara2019} clearly explains that ``a supervised machine learning tool is only as good as the data used for its training''. Since spambot detection is a rapidly-evolving field, we are all involved in the quest for up-to-date datasets. By playing with the parameters of the evolutionary algorithm here proposed, we advocate the capability to create a huge variety of fresh data, to re-train and fine-tune existing detection mechanisms.




\section{Related Work}
\label{sec:RW}
Although representing an effective heuristic to solve complex optimization problems~\cite{Mitchell:1998}, genetic algorithms usually require a string-based genetic encoding of information to be applied. This requirement severely limited their applicability. 
Remarkably, in recent years, we assisted to the proliferation of many studies on modeling and analyzing online behaviors. 
A stream of research focused on specific behavioral analytics tasks, such as detecting specific behavioral patterns~\cite{salathe2013dynamics,bagci2015random,Kumar:2017}, predicting future behaviors~\cite{zhou2012social,Li:2014:PHA}, and detecting anomalous ones~\cite{zafarani201510,cresci2016dna,ferrara2017,cresci2017social,cai2017detecting,yuan2017spectrum}. Others instead achieved more general results. In~\cite{Jeong:2017} authors showed that individuals have persistent and distinct online inter-event time distributions, while~\cite{dsaa2017} focused on modeling human tweeting behaviors, showing that such behaviors are very diverse and heterogeneous, although far from being random.
One result achieved in behavioral analytics is the possibility to encode the behavioral information of an account in a DNA-like string of characters~\cite{cresci2016dna,dsaa2017}.
The characterization of the behavior of both legitimate accounts and spambots through this \textit{digital DNA} modeling technique, coupled with the capability to carry out evolutionary simulations by means of genetic algorithms, opens up the unprecedented opportunity to quantitatively study and experiment with spambot evolutions and motivates our research.

Meanwhile, progress has been made towards the detection of malicious accounts (e.g., fakes, bots, spammers). As such accounts put in place complex mechanisms to evade existing detection systems, scholars tried to keep pace by proposing powerful techniques based on profile-~\cite{cresci2015,zafarani201510,badri2016uncovering}, posting-~\cite{wu2015social,Giatsoglou2015,chavoshi2016debot,badri2016uncovering,cresci2016dna,cresci2017social}, and network-characteristics~\cite{yang2014uncovering,yu2014,wu2015social,badri2016uncovering,yuan2017spectrum,WangGF17,liu2017holoscope} of the accounts. However, until now, new detection systems have been developed only as a consequence of spambot evolutions~\cite{Cresci2017}. In fact, no work has ever been done towards studying, and possibly anticipating, such evolutions. In other words, malicious accounts detection has always been tackled with a \textit{reactive} approach, which is in contrast with the novel \textit{proactive} approach envisaged in this research.

\section{Conclusions}
\label{sec:concl}
We presented the first exploratory study to carry out a quantitative analysis of spambot evolutions.
Specifically, riding the wave of the adversarial learning line of research, we first designed a novel genetic algorithm for simulating spambot behavioral evolutions. 
Then, we evaluated the extent to which the evolved spambots are capable of evading 3 state-of-the-art detection systems, based on evaluating the accounts' behavior and the content of posts. Testing the first system, results showed that as much as 79\% of the evolved spambots evade detection. Additionally, we also investigated the characteristics of the evolved spambots, highlighting distinctive features (e.g., entropy within digital DNA sequences) that would allow to distinguish them from legitimate accounts. Considering these features in current detection systems based on behavioral characteristics of the accounts would make them more robust against possible future spambot evolutions.

Further experimentation with the proposed approach could lead to new interesting results not only in bot design (e.g., chatbots), but also and foremost in the development of spambot detection systems. Indeed, until now, researchers had to \emph{react} to bot evolutions. However, for the first time since the advent of OSNs, there is the concrete chance to \emph{proactively} tackle the challenging task of spambot detection. Although here instantiated for a specific modeling and detection technique based on the behavior of the accounts, we thus argue that the proposed proactive approach will  provide the scientific community with the possibility to experiment with and simulate future spambot evolutions, 
substantially raising the bar for spambot developers.

\balance
\bibliographystyle{ACM-Reference-Format}
\bibliography{references}

\end{document}